\documentclass{ws-procs975x65}
\usepackage{rotating,cite,graphicx}
\newcommand{\IS}[4]{
\renewcommand{\arraystretch}{0.5}
\begin{array}{r}\mbox{\scriptsize #1} \\ \mbox{\scriptsize #2}\end{array}\mbox{#3}_{#4}
\renewcommand{\arraystretch}{1.0}
}
\newcommand{\MST}[1]{\scriptscriptstyle{#1}}
\begin{document}

\title{Nuclear Supersymmetry: New Tests and Extensions}

\author{A. Frank,$^{1,2)}$ J. Barea$^{1)}$ and R. Bijker$^{1)}$}

\address{$^{1)}$Instituto de Ciencias Nucleares, UNAM, Circuito Ext., C.U., A.P. 70-543, 04510 M\'exico, D.F. \\
$^{2)}$Centro de Ciencias F\'\i sicas,
Universidad Nacional Aut\'onoma de M\'exico,\\
Apartado Postal 139-B, 62251 Cuernavaca, Morelos, M\'exico}


\maketitle

\abstracts{
Extensions of nuclear supersymmetry are discussed, together with a proposal for new, more stringent and
precise tests that probe the susy classification and specific two-particle correlations among supersymmetric
partners.  The combination of these theoretical and experimental studies may play a unifying role in
nuclear phenomena.}

\section{Introduction}
 
Nuclear supersymmetry (n-susy), first proposed by Franco Iachello more than two
decades ago\cite{uno}, is a composite-particle phenomenon linking the
properties of bosonic and fermionic systems, framed in the context of the
Interacting Boson Model of nuclear structure\cite{dos}.  Composite particles,
such as the $\alpha$-particle
are known to behave as approximate bosons.  As He atoms they become superfluid
at low temperatures, an under certain conditions can also form Bose-Einstein
condensates.  At higher densities (or temperatures) the constituent fermions  begin to
be felt and the Pauli principle sets in.  Odd-particle composite systems, on
the other hand, behave as approximate fermions, which in the case of the IBFM
are treated as a combination of bosons and an (ideal) fermion.  In contrast to
the theoretical construct of supersymmetric particle physics, where susy is
postulated as a generalization of the Lorentz-Poincar\'e invariance at a
fundamental level, n-susy has been subject to experimental verification\cite{tres}.
Nuclear supersymmetry should not be confused with fundamental
susy, which predicts the existence of supersymmetric particles, such as the
photino and the selectron, for which no evidence has yet been found.  If such
particles exist, however, susy must be strongly broken, since large mass
differences must exist among superpartners, or otherwise they would have been
already detected.  Competing susy models give rise to diverse mass predictions
and are the basis for current superstring and brane theories\cite{cuatro}.  

Nuclear supersymmetry, on the other hand, is a theory that establishes precise
links among the spectroscopic properties of certain neighboring nuclei. 
Even-even and odd-odd nuclei are composite bosonic systems, while odd-A nuclei
are fermionic.  It is in this context that n-susy provides a theoretical
framework where bosons and fermions are treated as members of the same
supermultiplet\cite{cinco}.  the mass difference between superpartners is thus
of the order of 1/A, but the theory goes much further and treats the excitation
spectra and transition intensities of the  different systems as arising from a
single Hamiltonian and a single set of transition operators. 
Originally framed as a symmetry among pairs of nuclei\cite{uno,dos,cinco}, it
was subsequently extended to nuclear quartets or ``magic squares'', where
odd-odd nuclei could be incorporated in a natural way\cite{seis}.  Evidence
for the existence of n-susy (albeit possibly significantly broken) grew over
the years, specially for the quartet of Fig. 1, but only recently  more  systematic evidence was
found.  This was achieved  by means of one-nucleon transfer reaction experiments leading to the
odd-odd nucleus $^{196}$Au, which, together with the other members of the susy
quartet ($^{194}$Pt, $^{195}$Au and $^{195}$Pt) is considered to be the best
example of n-susy in nature\cite{seis,siete,ocho,nueve}.  We should point out,
however, that while these experiments provided the first complete energy
classification for $^{196}$Au (which was found to be consistent with the
theoretical predictions\cite{seis,siete,ocho,nueve}), the reactions involved
($^{197}$Au$(\vec d,t)$, $^{197}$Au$(p,d)$ and $^{198}$Hg$(\vec d,\alpha)$) did
not actually test directly the supersymmetric wave functions, as we shall
discuss below.  Furthermore, whereas these new measurements are very exciting, the
dynamical susy framework is so restrictive that there was little hope that
other quartets could be found and used to verify the theory\cite{seis,siete,ocho,nueve}. 
The purpose of this paper is two-fold.  On the one hand we report on an ongoing
investigation of one- and two-nucleon transfer reactions\cite{diez} in the
Pt-Au region that will more directly analyze the supersymmetric wave functions
and measure new correlations which have not been tested up to now.  On the
other hand we discuss some ideas put forward several years ago, which
question the need for dynamical symmetries in order for n-susy to exist\cite{once}.
  We thus propose a more general theoretical framework for nuclear
supersymmetry.  The combination of such a generalized  form of supersymmetry and
the  transfer experiments now  being carried out\cite{doce}, could provide
remarkable new correlations and a unifying theme in  nuclear structure
physics.

\section{New experiments}

\begin{figure}[t]
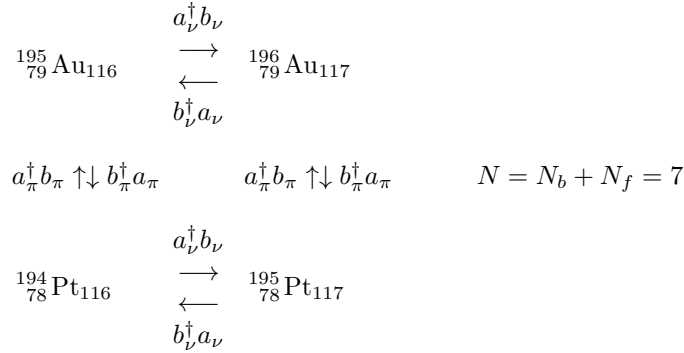


\begin{eqnarray}
\begin{array}{lclc}
\IS{195}{79}{Au}{116} & 
\begin{array}{cc}
a^\dagger_\nu b_\nu &  \\ 
\longrightarrow &  \\ 
\longleftarrow &  \\ 
b^\dagger_\nu a_\nu & 
\end{array}
& \IS{196}{79}{Au}{117} &  \\ 
&  &  &  \\ 
a^\dagger_\pi b_\pi \uparrow\downarrow b^\dagger_\pi a_\pi &  & 
a^\dagger_\pi b_\pi \uparrow\downarrow b^\dagger_\pi a_\pi & \hspace{1cm} N
= N_b + N_f = 7 \\ 
&  &  &  \\ 
\IS{194}{78}{Pt}{116} & 
\begin{array}{cc}
a^\dagger_\nu b_\nu &  \\ 
\longrightarrow &  \\ 
\longleftarrow &  \\ 
b^\dagger_\nu a_\nu & 
\end{array}
& \IS{195}{78}{Pt}{117} & 
\end{array}
\nonumber
\end{eqnarray}%

\caption{The magic quartet of nuclei.  The one-nucleon transfer reactions among them are indicated schematically, with $a^\dagger$ corresponding to a fermion and $b^\dagger$ to a boson of the indicated particles.}  
\end{figure}

\noindent The quartet of nuclei of Fig. 1 was classified by means of the $U_\nu(6/12)\times U_\pi(6/4)$ dynamical
supersymmetry, obtained by combining the $SO^{BF}(6)$ symmetry limit for the odd neutron ($^{195}$Pt) and the
$Spin(6)$ symmetry limit for the odd proton ($^{195}$Au)\cite{seis}.  
The excitation spectra of the nuclei $^{194}$Pt, $^{195}$Au and $^{195}$Pt was used to determine the
Hamiltonian and subsequently the spectra of the odd-odd partner $^{196}$Au was predicted, for which at the
time little or no experimental data was available. One should note, however, that the great majority of tests
carried out for the supersymmetric framework have involved one-nucleon transfer experiments 
leading  to the nuclei in figure 1 through reactions coming from outside the quartet, such as
$^{197}$Au$(\vec d, t)^{196}$Au and $^{196}$Pt$(\vec d,t)^{195}$Pt that, in first approximation, are
formulated using a transfer operator of the form $a^\dagger_\nu$.  The latter reactions are useful to measure
energies, angular momenta and parity of the residual nucleus and in principle provide information about the
systems wave functions.  However, they cannot test correlations present in the quartet's wave functions and
thus in the susy classification scheme as is the case for one-nucleon transfer reactions inside the
supermultiplet. These reactions do provide a direct test of the fermionic sector of the
graded Lie Algebras $U_\nu(6/12)$ and $U_\pi (6/4)$.
These operators are related to the nondiagonal elements of the product:
\begin{eqnarray}
U_\nu (6/12)\otimes U_\pi (6/4): \left(\begin{array}{ll}
b^\dagger_\nu b_\nu & b^\dagger_\nu a_\nu \\ &  \\ a^\dagger_\nu b_\nu & a^\dagger_\nu a_\nu \end{array} \right)
\oplus \left(\begin{array}{ll}
b^\dagger_\pi b_\pi & b^\dagger_\pi a_\pi \\  & \\
a^\dagger_\pi b_\pi & a^\dagger_\pi a_\pi \end{array}\right).
\end{eqnarray}


New experimental facilities and detection techniques\cite{siete,ocho,nueve} offer a unique opportunity for
analyzing the supersymmetry classification in greater detail\cite{doce}.  In reference\cite{trece} we pointed
out a symmetry route for the theoretical analysis of such reactions, via the use of tensor operators of the
algebras and superalgebras.  An alternative route is the use of a semi-microscopic approach where projection
techniques starting from the original  nucleon pairs lead to specific forms for  the
operators\cite{catorce,quince} which, however, are only strictly  valid in the generalized seniority
regime\cite{dieciseis}.  The former and latter routes may be related by a consistent-operator approach, where
the Hamiltonian exchange operators are made to be consistent with the one-nucleon transfer operator  
implying that the exchange term in the boson-fermion Hamiltonian can be viewed as an internal exchange
reaction  among the nucleon and the nucleon pairs. 

In addition to these experiments, we are also exploring the possibility of testing susy through new transfer
reactions.  The two-nucleon transfer $(\alpha, \vec d)$ reaction probes $n-p$ correlations in the nuclear
wave function and constitutes a very stringent test of the supersymmetry classification.  Note also that the
$^{194}$Pt$(\alpha, \vec d)^{196}$Au reaction corresponds to a combination of the single-nucleon transfers
going either through $^{195}$Pt or through $^{195}$Au, and that the corresponding operator is thus a product
of the fermionic components in equation (1),  as schematically indicated in

\begin{figure}[t]
\epsfxsize=15pc 
\begin{center} \begin{turn}{45} \epsfbox{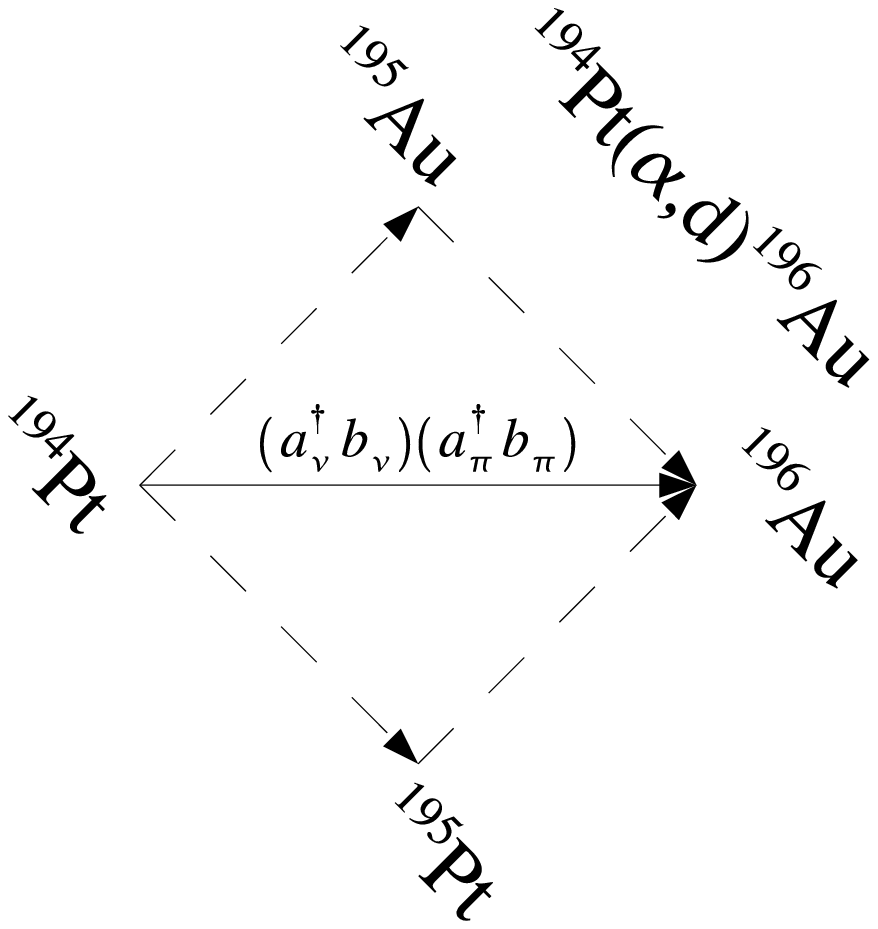} \end{turn} \end{center} 
\vspace*{-2cm}
\caption{The ``diagonal'' $(\alpha, d)$ reaction within the susy quartet.}  
\end{figure}

\noindent Fig. (2).  Likewise, the reaction $^{195}$Pt$(^3He,t)^{195}$Au, is again expressible in terms of
the superalgebra fermionic operators in (1) and in this case is associated to the beta-decay
operator\cite{dieciocho}.  These reactions and their relation to single-nucleon transfer experiments raise
the exciting possibility of testing direct correlations among transfer reaction spectroscopic factors in
different nuclei, predicted by the supersymmetric classification of the magic quartet.  A preliminary report
on these analyses is  presented in Ref.~\cite{diez}.

\section{Susy without Dynamical Symmetry}

\begin{figure}[t]
\epsfxsize=15pc 
\begin{center}
\begin{tabular}{lr}
\epsfbox{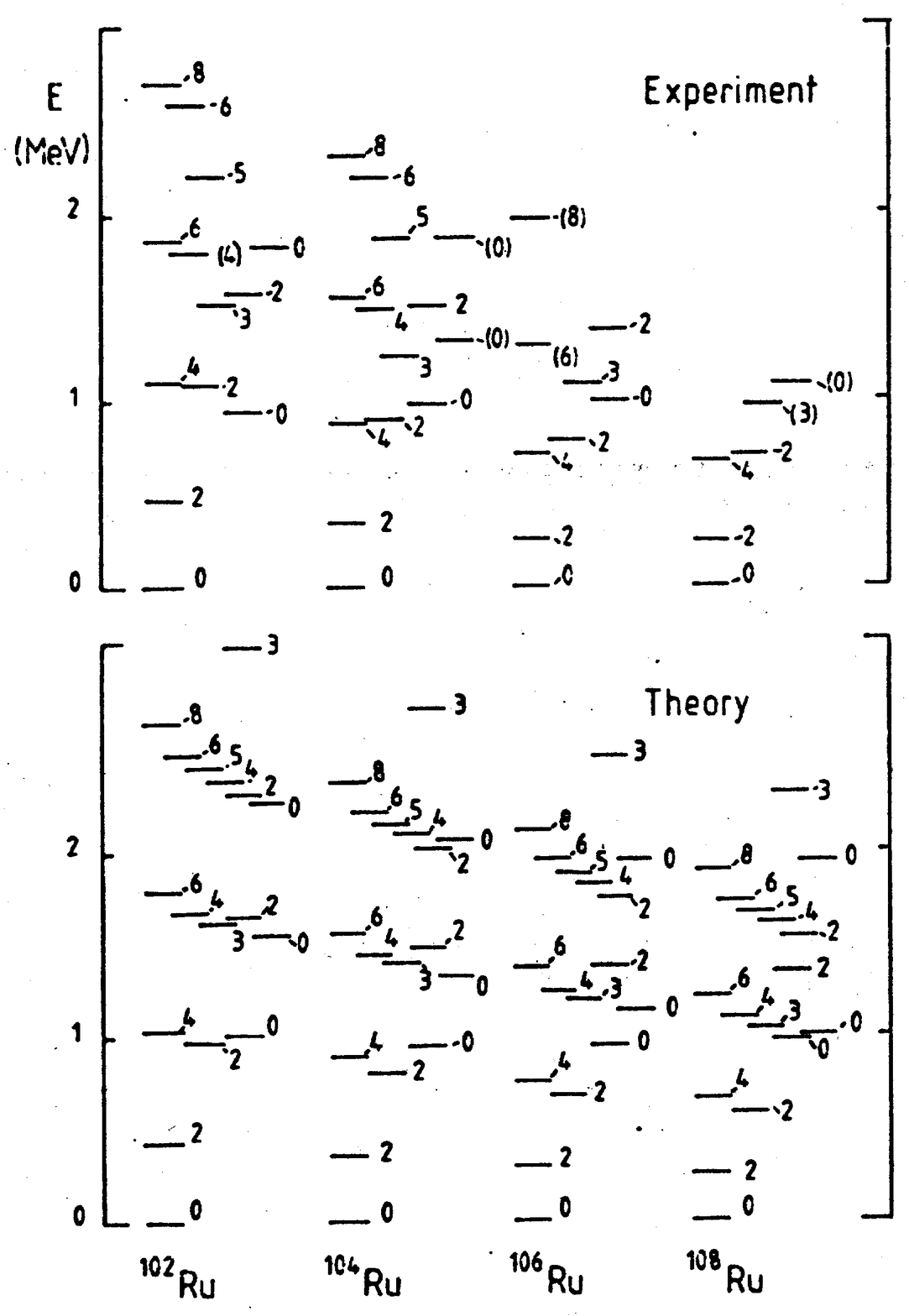} & \epsfbox{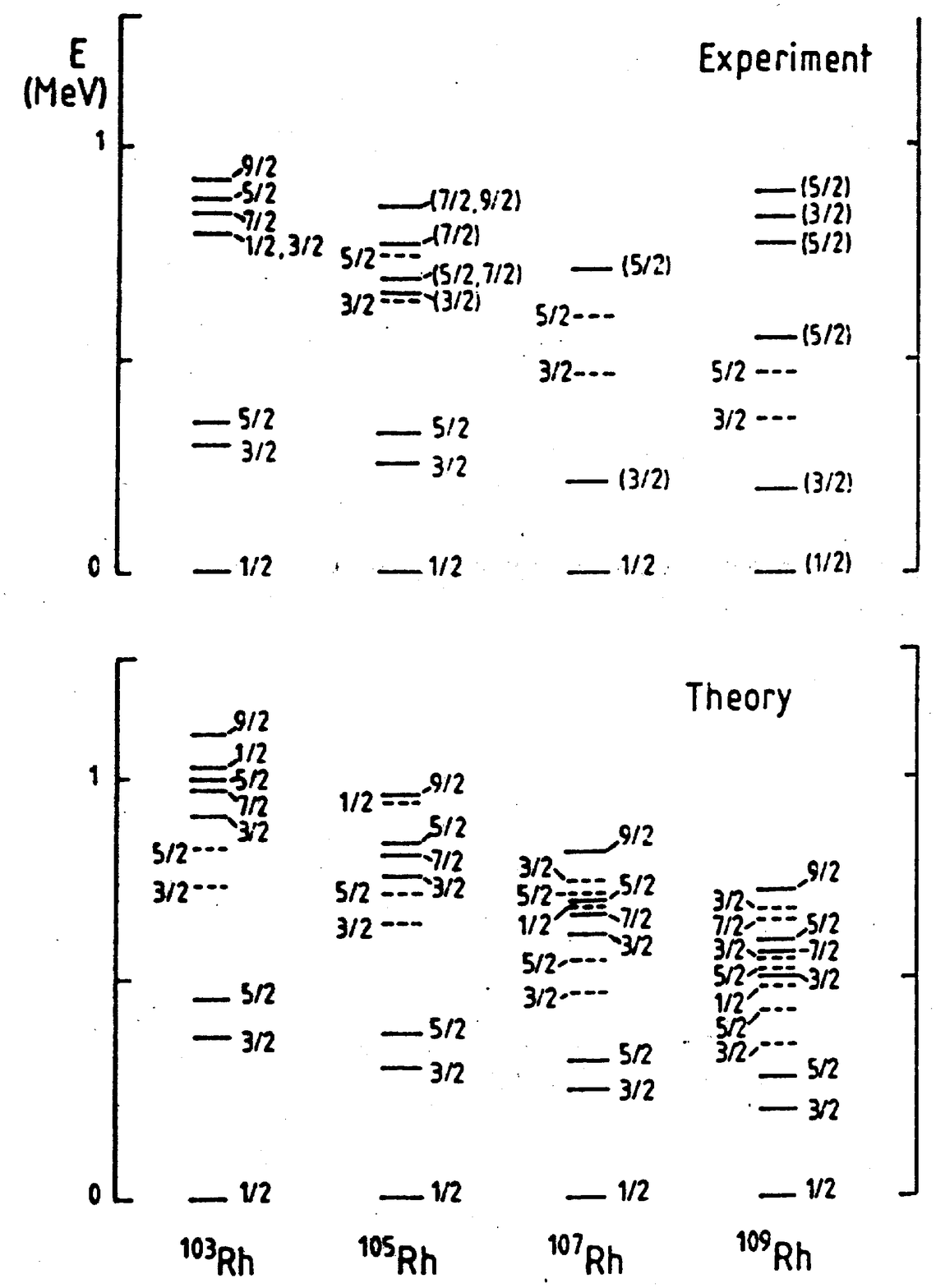} 
\end{tabular} 
\end{center}
\caption{Experimental and calculated positive-parity states in $^{102-108}Ru$ and
negative-parity states in $^{103-109}Rh$. Taken from \cite{once}.}  
\end{figure}

The concept of dynamical algebra (not to be confused with that of dynamical symmetry) implies a
generalization of the concept of symmetry algebra.  The latter is defined as follows: $G$ is the dynamical
algebra of a system if \underbar{all} physical states considered belong to a single irreducible representation (IR) of
$G$. (In a symmetry algebra, in contrast, each set of degenerate states of the system is associated to an
IR).  The best known examples of a dynamical algebra are perhaps  $SO(4,2)$ for the hydrogen atom and the $U(6)$ IBM algebra for
even-even nuclei.  A consequence of having a dynamical algebra associated to a system is that all sates can
be reached using the algebra's generators or, equivalently, all physical operators can be expressed in terms
of these operators\cite{nueve}.  Naturally, the same Hamiltonian and the same transition operators are
employed for all states in the system.  To further clarify this point, it is certainly true that a single $H$
and a single set of  operators are associated to a given even-even nucleus in the IBM framework, expressed in
terms of the $U(6)$ (dynamical algebra) generators.  It doesn't matter whether this Hamiltonian can be
expressed or not in terms of the generators of a single chain of groups (a dynamical symmetry).  

In the same fashion, if we now consider $U(6/12)$ to be the dynamical algebra for the pair of nuclei
$^{194}$Pt-$^{195}$Pt, it follows that the same $H$ and operators (including in this case the transfer
operators that connect states in the different nuclei) should apply to all states.  It also follows that no
restriction should be imposed on the form of $H$, except that it must be a function of the generators of
$U(6/12)$ (the enveloping space associated to it).  It should be clear that the concept of supersymmetry does
not require the existence of a particular dynamical symmetry.  Extending these ideas to the $\nu-\pi$ space
of IBM-2 we can say that susy is equivalent to requiring that a product of the form  \begin{equation} U_\nu
(6/\Omega) \otimes U_\pi (6/\Omega^\prime) \end{equation} plays the role of dynamical (super) algebra for a
quartet of even-even, even-odd, odd-even and odd-odd nuclei.  Having said that, it should be stated that the
quartet \underbar{dynamical} susy of references~\cite{seis,siete,ocho,nueve} has the distinct advantage of
immediately suggesting the form of the quartet's Hamiltonian and operators, while the general statement made
above does not provide a general recipe.  For some particular cases, however, this can be done in a
straightforward way.  In reference~\cite{once}, for example, the $U(6/12)$ supersymmetry (without
imposing one of the three dynamical IBM symmetries) was successfully tested for the Ru and Rh isotopes.  In
that case a combination of $U^{B+F}(5)$ and $SO^{B+F}(6)$ symmetries was shown to give an excellent
description of the data, as shown in Figs. 3 and 4.  The $U(6/12)$ case is simple because, using a pseudospin
decomposition, there are isomorphic $U(6)$ algebras for the bosons and the fermion and any combination of the
three dynamical IBM algebras can be considered\cite{diecinueve}  \begin{eqnarray} & & U (6/12) \quad j =
1/2,  3/2, 5/2 \nonumber \\ & & \tilde l = 0, 2 \qquad \tilde s = 1/2 \nonumber \\ & & G^{BF}_l \equiv G^B_l
+ G^F_{\tilde l} ~~ ,  \end{eqnarray} and an arbitrary interaction expressed in terms of $G^{BF}$ implies
explicit correlations between the boson-boson  and boson-fermion interactions\cite{diecinueve}.

An immediate consequence of this proposal is that it opens up the possibility of testing susy in other
nuclear regions, since dynamical symmetries are very scarce and have severely limited the study of nuclear
supersymmetry. 

\section{Generic Susy}

\begin{figure}[t]
\epsfxsize=15pc 
\begin{center}
\epsfbox{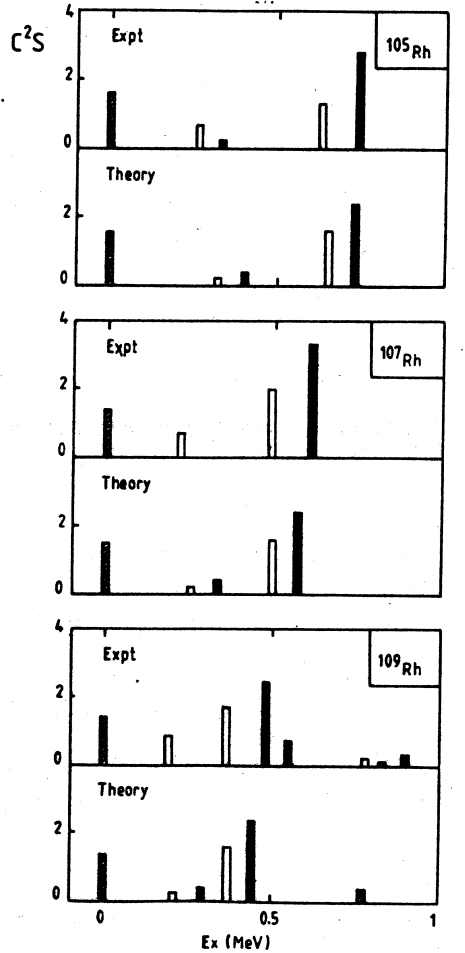}
\end{center}
\caption{Experimental and calculated spectroscopic factors in Rh isotopes. Taken from \cite{once}.}  
\end{figure}

We have recently initiated a renewed search for supersymmetry in nuclei\cite{doce,diecinueve}.  We have yet
to discover a general mechanism to generate all appropriate operators in the general case, but a set of
guiding rules are the following:

1)  The Hamiltonian should describe the members of the doublet or quartet. 

2)  The boson-Hamiltonian, plus the single-particle orbits, should essentially determine the boson-fermion
interaction, for both the odd-proton and odd-neutron nuclei. 

3)  The combination of the previous three should give a prediction for the odd-odd Hamiltonian, and thus about
the $p-n$ interaction.

Although the analysis is not concluded, our preliminary results for the $W$ and $Hf$ nuclei are quite
encouraging\cite{diecinueve}. The first calculation involves a mixture of $SU(3)$ and $SO(6)$ symmetries in
$U(6/12)$. This calculation employs the Q-consistent formalism and a comparison between the experimental and
calculated $BE(2)$ transitions and quadrupole moments is shown in tables 1 and 2. The agreement is
very good, except for one transition in $^{183}W$. We also show an example of generic susy in U(6/4). It
corresponds to the supermultiplet composed of $^{174}Hf$ and $^{173}Hf$. In this case the hamiltonian uses a
combination of Casimir operators of the $U(5)$ and $SO(6)$ groups and of their subgroups. In figure 5 we compare
the experimental and calculated level energies in these two nuclei\cite{diecinueve}.

\begin{table}[b]
\renewcommand{\arraystretch}{1.5}
\begin{minipage}[t]{12.8pc}
\begin{tabular}{ccc}
\hline 
\multicolumn{3}{c}{$B(E2)$ ($e^{2}b^{2}$) and $Q$ ($eb$) in $^{182}$W}\\
$J_{i}^{\pi }\rightarrow J_{f}^{\pi }$&
Exp.&
Calc.\\
\hline 
$2_{1}^{+}\rightarrow 0_{1}^{+}$&
0.839(18)&
0.8422\\
$4_{1}^{+}\rightarrow 2_{1}^{+}$&
1.201(61)&
1.1877\\
$6_{1}^{+}\rightarrow 4_{1}^{+}$&
1.225(135)&
1.2777\\
$2_{2}^{+}\rightarrow 0_{1}^{+}$&
0.021(1)&
0.0040\\
$2_{2}^{+}\rightarrow 2_{1}^{+}$&
0.041(1)&
0.0072\\
$2_{2}^{+}\rightarrow 4_{1}^{+}$&
0.00021(1)&
0.0006\\
$2_{3}^{+}\rightarrow 0_{1}^{+}$&
0.006(1)&
0.0000\\
$2_{3}^{+}\rightarrow 0_{2}^{+}$&
1.225(368)&
0.6840\\
$2_{3}^{+}\rightarrow 2_{1}^{+}$&
0.0039(5)&
0.0001\\
&
&
\\
\hline 
$Q$&
Exp.&
Calc.\\
\hline
$2_{1}^{+}$&
$-2.00_{-0.08}^{+0.04}$&
-1.86\\
$2_{2}^{+}$&
$1.94_{-0.04}^{+0.10}$&
1.61\\
\hline
\end{tabular}
\caption{Experimental and calculated reduced transition probabilities and
qua\-dru\-po\-le mo\-ments in $^{182}W$.
Taken from \cite{diecinueve}.}
\end{minipage}
\hspace{2cm}
\begin{minipage}[t]{12pc}
\begin{tabular}{ccc}
\hline 
\multicolumn{3}{c}{$B(E2)$ ($e^{2}b^{2}$) in $^{183}$W}\\
$J_{i}^{\pi }\rightarrow J_{f}^{\pi }$&
Exp.&
Calc.\\
\hline 
$\frac{3}{2}_{\MST 1}^{-}\rightarrow \frac{1}{2}_{\MST 1}^{-}$&
0.938(62)&
0.603\\
$\frac{5}{2}_{\MST 1}^{-}\rightarrow \frac{1}{2}_{\MST 1}^{-}$&
0.68(4)&
0.603\\
$\frac{5}{2}_{\MST 1}^{-}\rightarrow \frac{3}{2}_{\MST 1}^{-}$&
0.20(3)&
0.173\\
$\frac{13}{2}_{\MST 1}^{-}\rightarrow \frac{9}{2}_{\MST 1}^{-}$&
1.1(3)&
0.915\\
$\frac{17}{2}_{\MST 1}^{-}\rightarrow \frac{13}{2}_{\MST 1}^{-}$&
0.89(12)&
0.925\\
$\frac{3}{2}_{\MST *}^{-}\rightarrow \frac{1}{2}_{\MST 1}^{-}$&
0.005(2)&
0.000\\
$\frac{3}{2}_{\MST *}^{-}\rightarrow \frac{3}{2}_{\MST 1}^{-}$&
0.10(4)&
0.010\\
$\frac{3}{2}_{\MST *}^{-}\rightarrow \frac{5}{2}_{\MST 1}^{-}$&
0.012(5)&
0.023\\
$\frac{5}{2}_{\MST *}^{-}\rightarrow \frac{1}{2}_{\MST 1}^{-}$&
0.082(9)&
0.011\\
$\frac{5}{2}_{\MST *}^{-}\rightarrow \frac{3}{2}_{\MST 1}^{-}$&
0.001(1)&
0.001\\
$\frac{5}{2}_{\MST *}^{-}\rightarrow \frac{5}{2}_{\MST 1}^{-}$&
0.027(6)&
0.004\\
$\frac{5}{2}_{\MST *}^{-}\rightarrow \frac{7}{2}_{\MST 1}^{-}$&
0.43(2)&
0.001\\
$\frac{5}{2}_{\MST *}^{-}\rightarrow \frac{3}{2}_{\MST 2}^{-}$&
1.30(18)&
0.860\\
\hline
\end{tabular}
\caption{Experimental and calculated reduced transition probabilities in $^{183}W$.
Taken from \cite{diecinueve}.}
\end{minipage}
\renewcommand{\arraystretch}{1}
\end{table}

\begin{figure}[t]
\epsfxsize=20pc 
\begin{center}
\epsfbox{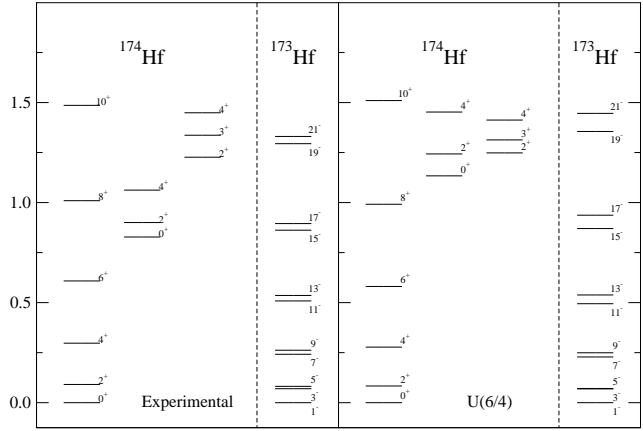} 
\end{center}
\caption{Experimental and calculated positive-parity states in $^{174}Hf$ and
negative-parity states in $^{173}Hf$.The energy scale is in MeV. Taken from \cite{diecinueve}.}  
\end{figure}

One of our main interests is to apply the generic form of supersymmetry to the Pt-Au region and compare the
results with the traditional scheme, particularly for the new transfer experiments\cite{doce}.  In addition,
we expect to find other examples of quartet supersymmetric behavior, once the constraints set by dynamical
symmetry are dropped. 

We continue to search for a more general theoretical framework that can encompass the particular cases that
we can solve at this point.  

We conclude by proposing that nuclear susy may be a more general phenomenon than was previously realized and
that may yet play an important, unifying role in nuclei.

\section{Dedication:}

\begin{figure}[t]
\epsfxsize=20pc 
\begin{center}
\epsfbox{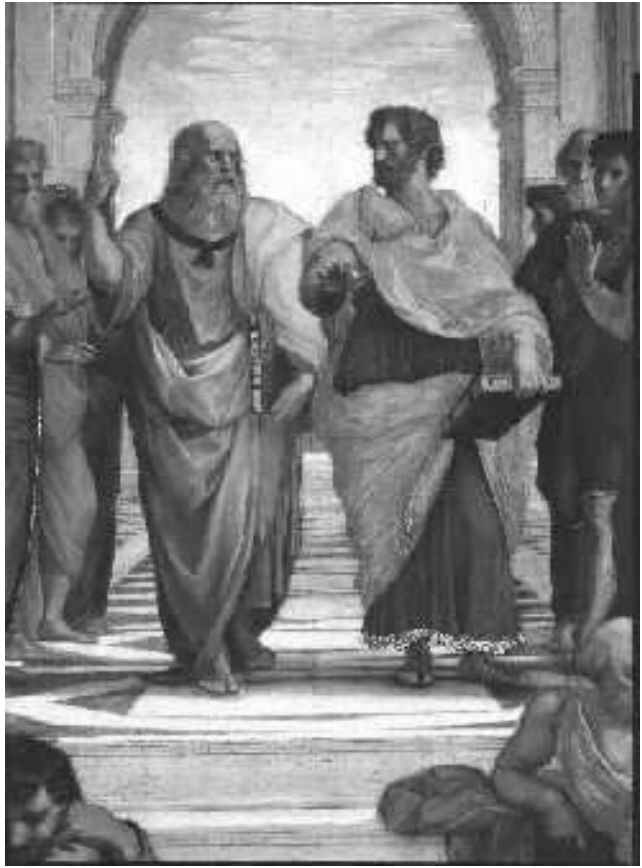} 
\end{center}
\caption{Detail of ``The School of Athens'' (Plato on the left and Aristoteles on the right), by Rafael.}  
\end{figure}

We dedicate this paper to Franco Iachello, who has managed to uniquely combine the Platonic ideal of symmetry
with the down-to-earth Aristotelic ability to recognize these patterns in Nature.

\section*{Acknowledgments}
We are grateful to P. Van Isacker, G. Graw, J. Jolie, J. 
Arias and C. Alonso for their collaboration and much inspiration.  This paper was supported in part by Conacyt, Mexico.


\begin{thebibliography}{99}
\bibitem{uno}F. Iachello, \Journal{\em Phys. Rev. Lett.}{44}{772}{1982}.

\bibitem{dos}F. Iachello and P. Van Isacker, {\em ``The Interacting Boson-Fermion Model''}, Cambridge University Press, Cambridge, (1991). 

\bibitem{tres}A. Metz, {\it et al}, \Journal{\em Phys. Rev. Lett.}{83}{1542}{1999} and references therein. 

\bibitem{cuatro}A. Kostelecky and D.K. Campbell, Eds., {\em ``Supersymmetry in Physics''}, North Holland,
Amsterdam (1984); S. Weinberg, {\em ``The quantum theory of fields: Supersymemtry''}, Cambridge (2000).

\bibitem{cinco}A.B. Balantekin, I. Bars and F. Iachello, \Journal{{\em Nucl. Phys.} A}{370}{284}{1981}; A.B.
Balantekin, I. Bars, R. Bijker and F. Iachello, \Journal{{\em Phys. Rev.} C}{27}{1761}{1983}; R. Bijker, Ph.D. Thesis, (1984); A.
Mauthofer {\it et al}, \Journal{{\em Phys. Rev.} C}{39}{1111}{1989}. 

\bibitem{seis}P. Van Isacker, J. Jolie, K. Heyde and a. Frank, \Journal{Phys. Rev. Lett.}{54}{653}{1985}; 

\bibitem{siete}D.D. Warner, R.F. Casten and A. Frank, \Journal{{Phys. Lett.} B}{180}{207}{1986}. 

\bibitem{ocho}A. Metz {\it et al}, \Journal{{Phys. Rev.} C}{61}{064313}{2000}. 

\bibitem{nueve}A. Frank and P. Van Isacker, {\em ``Algebraic Methods in Molecular and Nuclear Structure Physics''}, Wiley, New York (1994). 

\bibitem{diez}R. Bijker, J. Barea and A. Frank, these proceedings. 

\bibitem{once}A. Frank, P. Van Isacker and D.D. Warner, \Journal{{Phys. Lett.} B}{197}{474}{1987}. 

\bibitem{doce}G. Graw, private communication. 

\bibitem{trece}J. Barea {\it et al}, \Journal{{Phys. Rev.} C}{64}{64313}{2001}. 

\bibitem{catorce}O. Scholten, Ph.D. Thesis (1980). 

\bibitem{quince}J. Barea, C.E. Alonso and J.M. Arias, \Journal{{Phys. Rev.} C}{65}{34328}{2002}. 

\bibitem{dieciseis}I. Talmi, {\em ``Simple Models for Complex Nuclei''}, Harwood, (1993). 


\bibitem{dieciocho}P. Navr\'atil alnd J. Dobes, \Journal{{Phys. Rev.} C}{37}{2126}{1988}. 

\bibitem{diecinueve}J. Barea, Ph.D. Thesis (2002). 




\end{thebibliography}
\end{document}